\documentclass[a4paper]{article}

\usepackage{bibtopic} % pour separer les articles des sites web dans la biblio, 
\usepackage{alltt}

\usepackage{url}

\usepackage{boxedminipage}
\usepackage{listings}
\usepackage{subfigure}

\newcommand{\eg}[0]{\textit{e.g.},~}

\usepackage{graphicx} 

\usepackage{relsize}

\usepackage{color}

\title{Views, Program Transformations,\\ and the Evolutivity Problem\\ in a Functional Language\footnote{This is the second version of the report initially entitled \emph{Views, Program Transformations, and the Evolutivity Problem}.}}
\author{Julien Cohen$^1$
     \& R\'emi Douence$^2$}

\date{}

\begin{document}
\setcounter{page}{1}
\maketitle
% Here is the place for your text...

{\small
\noindent
1: Universit\'e de Nantes -- LINA (UMR 6241, CNRS, Univ. Nantes, \'EMN)\\
2: INRIA -- ASCOLA team (\'EMN - INRIA - LINA)\\
}

We report on an experience to support multiple views of
programs to solve the tyranny of the dominant decomposition
in a functional setting.
We consider two possible architectures in Haskell for the
classical example of the expression
problem.
We show how the Haskell Refactorer can be used to transform
one view into the other, and the other way back. That transformation is
automated and we discuss how the Haskell Refactorer has been
adapted to be able to support this automated transformation.
Finally, we compare our implementation of views with some of
the literature.
%%%%%%%%%%%%%%%%%%%%%%%%%%%%%%%%%%%%%%%%%%%%%%%%%%%%%

\section{Introduction}%%%%%%%%%%%%%%%%%%%%%%%%%%%%%%%%%%%%%%%%%%%%%%%%%%%
Evolutivity is a major criteria of quality for
enterprise software.
Evolutivity is strongly related to the
design choices on the
software architectures.
However, it is generally impossible to find software
architectures that are evolutive with respect to all
concerns.
So, one of these concerns has to be
privileged %in any architecture
(section~\ref{sec-primary-decomposition}).
As shown by the solutions to the \emph{expression problem}~\cite{expPb},
there are many ways, often based on specific language features, to provide modular extensions which are
orthogonal to the main axis of decomposition of the
architecture
(section~\ref{sec-expression-problem}).
However, these solutions focus on extensions and generally break 
the regularity of the initial architecture, leading to a decrease in the maintainability %ease of maintenance 
(section~\ref{sec-problem}).
%
%
%This problem is illustrated by the
%so-called \emph{expression problem}~\cite{expPb} (described
%in section~\ref{sec-expression-problem}) in a context where
%only extensions are taken into account.
%
%Many solutions have been proposed for the expression
%problem, see for instance~\cite{Zenger-Odersky2005} which reviews some of them.
%
%These solutions allow to extend a program in a modular way,
%by adding new compilation units (classes, modules, aspects,
%mixins...).
%
%So, they tackle extensibility, but the original expression
%problem formulation also requires strong static type safety,
%separate compilation and no modification of the existing
%code.
%
%However evolution is not only extension, it is also
%maintenance, for instance to correct an error or adapt a behavior.
%
%. We do not always want to add new
%functionalities, but we often need to modify the existing
%ones, for instance to correct an error or to
%change a behavior (corrective, adaptive or perfective
%maintenance~\cite{maintenance76}).
%
%
This 
shows that the modular extensibility and
maintainability on orthogonal concerns are difficultly
supported at the language level.

%But, since the expression problem focuses on extension, it
%does not address corrective or adaptative maintenance: its
%solutions provide no means to modify or add some code
%without degrading the initial architecture
%(section~\ref{sec-problem}).
%

%For this kind of evolution, solutions to the expression
%problem have a major drawback: maintenance is non modular
%(section~\ref{sec-problem}).
%
%Indeed, the code to modify, which was initially contained in
%a single module (or another compilation unit), is
%spread over several modules after extensions have been
%implemented.
%
%This is not surprising since the expression problem is
%focused on extension and forbids modifications of the
%existing code: it is too restricted to cover maintenance
%evolutions. For this reason, we do not limit ourselves to
%the expression problem.
%, and the more general problem
%described in section~\ref{sec-evolutivity-problem}~\ref{sec-problem}.

Multiple views~\cite{Black2004} tackle the problem of modular
evolutivity with a program transformation approach instead of a programming language approach.
For a given application, the source code of several
equivalent architectures can be computed one from another, so that
the programmer who has to implement an evolution can choose
the architecture which his the most convenient for his
task. With proper tools, the implemented evolutions are
reflected in all the available architectures.

In this paper, we report on an experience of providing
support for multiple views for a functional language.
In the following, 
%
%First, 
we consider the classical example of a simple evaluator coming from the expression problem~\cite{expPb} and we illustrate how multiple views can provide 
modular extensions as well as modular changes on several orthogonal
axis (section~\ref{sec-transformation}).
Then, we propose an implementation of a transformation from one view to another. That transformation is based
on a refactoring tool
(section~\ref{sec-refactoring-haskell}).
Last, we discuss the work to make this
kind of tool usable for enterprise software
(section~\ref{sec-usage}) and we compare our experience to
other tools for multiple views (section~\ref{related-work}).

%illustrate our idea with a program transformation between
%two functionally equivalent architectures
%(section~\ref{sec-refactoring-haskell}).
%
%That particular transformation is based on a sequence of
%basic refactoring operations and is supported by a
%refactoring tool.
%
%We also discuss how that transformation is subject to
%evolutions when needed (section~\ref{sec-evolution}).

\section{Modularity and Evolution} %%%%%%%%%%%%%%%%%%%%%%%%%%%%%%%%%%%%%%
In this section, we illustrate the tyranny of the dominant
decomposition in a functional language setting with a simple example
(section~\ref{sec-primary-decomposition}),
we recall the definition of the expression problem
(section~\ref{sec-expression-problem}), which is closely
related to our problem,
and we focus on maintenance and show that it is not
well covered by the expression problem as extensions tend to degrade the initial structure (section~\ref{sec-problem}).

\subsection{Each Architecture privileges extensibility on a given axis}%%%%%%%%%%%%%%%
\label{sec-primary-decomposition}
When choosing a module structure for a given program, one
has to choose between several possibilities with different
advantages and disadvantages~\cite{Parnas1972}.
We illustrate this with two possible module structures for a
simple evaluator which have dual advantages and
disadvantages. This program is the same that is often used
to motivate the expression problem, here given in
Haskell.
%\footnote{The examples of Haskell programs do not rely
%  on specificities of that language. However, our
%  implementation is limited to Haskell.}

\newcommand{\pfun}{P_{\mathit{fun}}}
\newcommand{\pdata}{P_{\mathit{data}}}

% options pour le package listing
\lstset{basicstyle=\small, frame=trBL, frameround=tfff, language=haskell, basicstyle=\ttfamily}

%\begin{figure}[!htp]
\begin{alltt}
 data Expr = 
     Const Int
   | Add (Expr,Expr)

 eval (Const i)     = i
 eval (Add (e1,e2)) = eval e1 + eval e2

 toString (Const i)     = show i
 toString (Add (e1,e2)) = (toString e1) ++ "+" ++ (toString e2)
\end{alltt}
%\caption{Expression Data Type (in Haskell) and two associated functions}
%\label{fig-haskell-datatype}
%\end{figure}

%
The data type
\texttt{Expr} represents the expression language to be
evaluated. This data type has a constructor for literals 
(\texttt{Const} for integers) and another for an operator (\eg \texttt{Add}
represents the addition).
Two functions, for evaluating or printing expressions, are defined
by pattern matching: a case is provided for each constructor.

\begin{figure}[!htp]

\centering
\subfigure[Functional decomposition (program $\pfun$)]{
\includegraphics[scale=0.65]{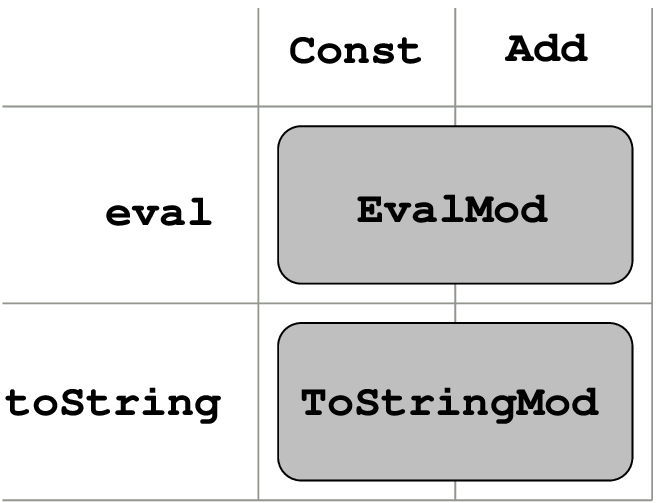}
%\caption{}
\label{fig-classical}
}
\hspace{3cm}
\subfigure[Constructor decomposition (program $\pdata$)]{
\includegraphics[scale=0.65]{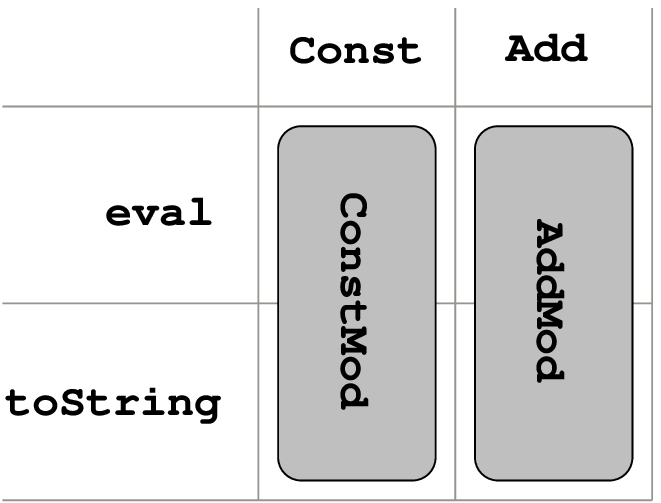} 
%\caption{}
\label{fig-dispatch}
}
\caption{Coverage of modules with respect to functions and data constructors}
\end{figure}

This code is modular with respect to functionalities
(because of the scope introduced by function
declarations). The modularity is better seen in
Figs.~\ref{fig-haskell-classical} and~\ref{fig-classical}
where modules are used to structure the program.

\begin{figure}[htp]
%\lstinputlisting{HASKELL/PFUN_WITHOUT_EXPORT/Expr.hs}
%\lstinputlisting{HASKELL/PFUN_WITHOUT_EXPORT/EvalMod.hs}
%\lstinputlisting{HASKELL/PFUN_WITHOUT_EXPORT/StringOfExprMod.hs}
%\lstinputlisting{HASKELL/PFUN_WITHOUT_EXPORT/Main.hs}
\begin{lstlisting}
module Expr  where

data Expr = 
     Const Int
   | Add (Expr,Expr)

e1 = Add (Add (Const 1,Const 2),Const 3)
e2 = Add (Add (Const 0,Const 1),Const 2)
\end{lstlisting}
\begin{lstlisting}
module EvalMod where
import Expr 

eval (Const i)     = i
eval (Add (e1,e2)) = eval e1 + eval e2
\end{lstlisting}
\begin{lstlisting}
module ToStringMod where
import Expr

toString (Const i)     = show i
toString (Add (e1,e2)) = (toString e1) ++ "+" ++ (toString e2)

\end{lstlisting}
\begin{lstlisting}
module Client where 
import Expr
import ToStringMod
import EvalMod

r1 = print (toString e1)
r2 = print (show (eval e1))
r3 = print (toString e2)
r4 = print (show (eval e2))
\end{lstlisting}
\caption{Functional decomposition in Haskell -- program $\pfun$}
\label{fig-haskell-classical}
\end{figure}

%
%The left-hand-side diagram of Figure~\ref{fig-classical} is the graph of dependencies between the modules. For instance, the module \texttt{EvalMod}
%depends on the data type defined in the
%module \texttt{Expr}.

%We see on this graph that \texttt{EvalMod} and
%\texttt{StringOfExprMod} are independent (there is no path
%from one to the other) which means that a change localized
%to one of them has no impact on the other (the evolutions in
%functionalities are modular).

Fig.~\ref{fig-classical} shows a matrix indexed on
constructors and functions where the modules of interest
have been pictured.
For instance, the module \texttt{EvalMod} deals with the two 
constructors but with only one function.

This program architecture makes it easy to modify an
existing function since the code to deal with is localized
by syntactic module boundaries.
It is also easy to add a new
function by adding a new module.
However, this code is not modular with respect to
data constructors. The code corresponding to a given constructor
(\eg \texttt{Add}) crosses module boundaries. So, when the
data type is extended and a new constructor (\eg
\texttt{Mult}) is introduced, each function module must be
modified in order to take into account the new
constructor.

\begin{figure}[htp]
\begin{lstlisting}
module Expr  where

data Expr = 
     Const Int
   | Add (Expr,Expr)

fold a c (Const i)     = c i
fold a c (Add (e1,e2)) = a (fold a c e1) (fold a c e2)

e1 = Add (Add (Const 1,Const 2),Const 3)
e2 = Add (Add (Const 0,Const 1),Const 2)
\end{lstlisting}
\begin{lstlisting}
module ConstMod (toString,eval) where

 toString x = show x
 
 eval x = x
 
\end{lstlisting}
\begin{lstlisting}
module AddMod (toString,eval) where

 toString x y = x ++ "+" ++ y
 
 eval x y = x + y
 
\end{lstlisting}
\begin{lstlisting}
module Client where 
import Expr
import ConstMod (eval,toString)
import AddMod (eval,toString)

toString x = fold AddMod.toString ConstMod.toString x

eval x = fold AddMod.eval ConstMod.eval x

r1 = print (Client.toString e1)
r2 = print (show (Client.eval e1))
r3 = print (Client.toString e2)
r4 = print (show (Client.eval e2))
\end{lstlisting}
\caption{Constructor decomposition in Haskell -- program $\pdata$}
\label{fig-haskell-dispatch}
\end{figure}

Figs.~\ref{fig-haskell-dispatch} and~\ref{fig-dispatch}
describe an alternate code architecture.
That structure gathers all the pieces of code
related to a given constructor into a single
module. 
For instance, the module \texttt{ConstMod} collects the
parts of the definition of \texttt{eval} and
\texttt{toString} for the \texttt{Const} case.
Fig.~\ref{fig-dispatch} pictures this architecture: modules
in the matrix do not cover functions anymore but
constructors.
%
%The Figure~\ref{fig-haskell-dispatch} details the
%corresponding code.  In particular, a function
%\texttt{dispatch} takes functions as parameters as well as
%and expression and it performs pattern matching for applying
%the right one.  In this context, a function such as
%\texttt{eval} is defined as the (partial) application of
%\texttt{dispatch} to the functions for evaluation.

This alternative code is modular with respect to
data constructors. 
Indeed, this program structure makes it easy (modular) to
add a new constructor (\eg \texttt{Mult} for a product): the
corresponding module is introduced (and \texttt{fold} is
extended with a new case). 
However, this code is not modular with respect to
functionalities: the code corresponding to a given function
(\eg \texttt{eval}) is spread in all constructor
modules. So, when a new function is
introduced, each module must be modified in order to take
 the implement the new function.

This illustrates the tyranny of the dominant decomposition in
action. Whatever primary program structure is chosen, some
extensions will not be modular.

\subsection{The Expression Problem}%%%%%%%%%%%%%%%%%%%%%%%%%%%%%%%%%%%%%%%%%%
\label{sec-expression-problem}

The problem we have described has been subject to many proposals
in the context of the so-called Expression Problem
(see~\cite{Zenger-Odersky2005} for a review of some
solutions). 
The expression problem tries to tackle modular
extensibility from a language point of view and imposes
constraints that are coherent for this point of view. These
constraints are the following~\cite{expPb}:

\begin{itemize}

\item The extension should come as a separate file/module and should not require to modify existing files/modules. 

\item The files/modules that were already in the program before the extension should not be recompiled.

\item The type system should be able to ensure that the extension is safe.

%\item COMPLETER / VERIFIER

\end{itemize}

Several works (for instance those listed
in~\cite{Zenger-Odersky2005}) show that specific features of the host language makes it possible to
design a program structure where it is modular to extend the
data type, and also modular to extend the functionalities. 
However, as we will see, these solutions share a drawback:
maintenance is not modular. 
Indeed, successive
evolutions tend to break the initial structure~\cite{BeladyLehman1971}.
This is
not taken into account by the expression problem.

%Moreover, we do not limit ourselves to a language approach (we will not introduce new language constructs but we will rely on program transformation tools), and we prefer to maintain at any time the ``right'' architecture (by changing the module boundary/definition) rather than limiting recompilation.
%are not interested in not recompiling unmodified modules (\emph{expliquer pourquoi cette contrainte est importante pour Wadler et pas pour nous}).

\subsection{Extension is only part of the problem}%%%%%%%%%%%%%%%%%%%%%%%%%%%
\label{sec-problem}

In order to illustrate the deviation from structural
regularity with referenced solutions to the expression
problem, we consider an incremental development scenario for
our example of application. 
%
%let us now make abstraction of the particular data
%type and functions of our example, and  for an abstract version of
%our example application.
%
 However, we abstract the code of our example and consider a
 data-type with two constructors \texttt{C1} and \texttt{C2}
 (instead of \texttt{Const} and \texttt{Add}), as well as
 two functions \texttt{f1} and \texttt{f2} (instead
 of \texttt{eval} and \texttt{toString}).

The initial program considered in this scenario has an
architecture focusing either on function extensibility or on data
extensibility, depending on a design choice.  These two
possible architectures are pictured in the following two
diagrams.\\

~\hfill\includegraphics[scale=0.8]{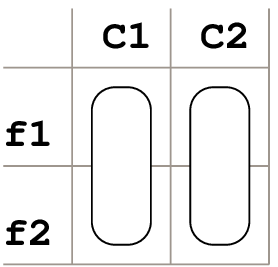} \hfill
\includegraphics[scale=0.8]{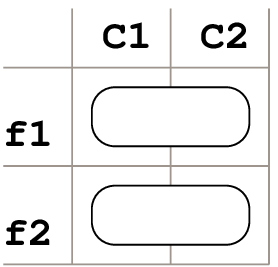}\hfill~\\

The left hand side diagram means that there is a
  module\footnote{We generalize the definition of module to:
  ``any modular entity of the programming language". For
  instance, a function definition is a modular entity. } for
  each constructor of the data type and the code of the
  functions is spread over these modules. This illustrates
  the situation in the architecture of
  Fig.~\ref{fig-haskell-dispatch} and also in the classical
  object approach (Composite design pattern).
The right hand side diagram means that there is a module for each
function and the code corresponding to a constructor
of the data type is spread over these modules. This
illustrates the situation of the classical functional
approach (Fig.~\ref{fig-haskell-classical}) and also in the Visitor design pattern.

\begin{figure}[htp]
\noindent\begin{boxedminipage}{\textwidth}
\medskip
\begin{enumerate}
\item Extension : Introduce a new constructor \texttt{C3} (for instance \texttt{Mult}).
\item Extension :  Introduce a new function \texttt{f3} (for instance \texttt{derive}).
\item Extension : Introduce a new constructor \texttt{C4} (for instance \texttt{Div}).
\item Extension : Introduce a new function \texttt{f4} (for instance \texttt{check\_div\_by\_zero}).
\item Maintenance : Modify the function \texttt{f1}.
\item Maintenance : Modify the data constructor \texttt{C1}.\\
\end{enumerate}
\end{boxedminipage}
\caption{Evolution scenario}
\label{fig-scenario-definition}
\end{figure}
\label{sec-scenario}

\begin{figure}[htp]
\centering
\subfigure[Progress with a language-level solution for two initial architectures.]{
\includegraphics[scale=0.6]{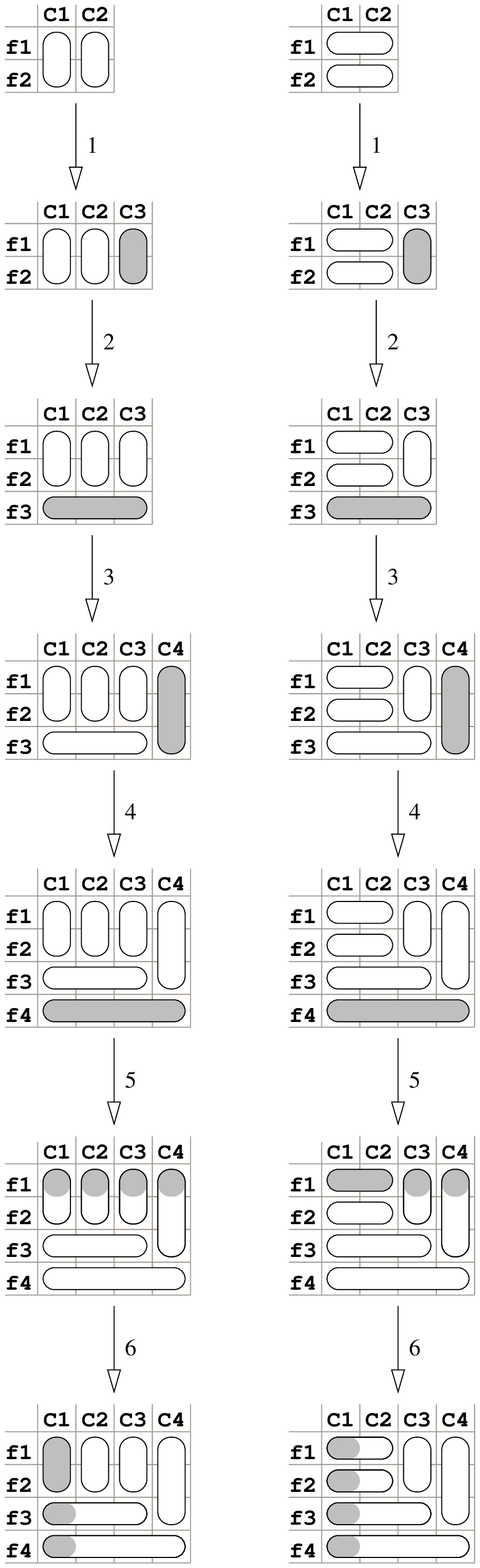}
\label{fig-current-solutions}
}
\hspace{1cm}
\subfigure[Progress with architecture transformations.]{
\includegraphics[scale=0.6]{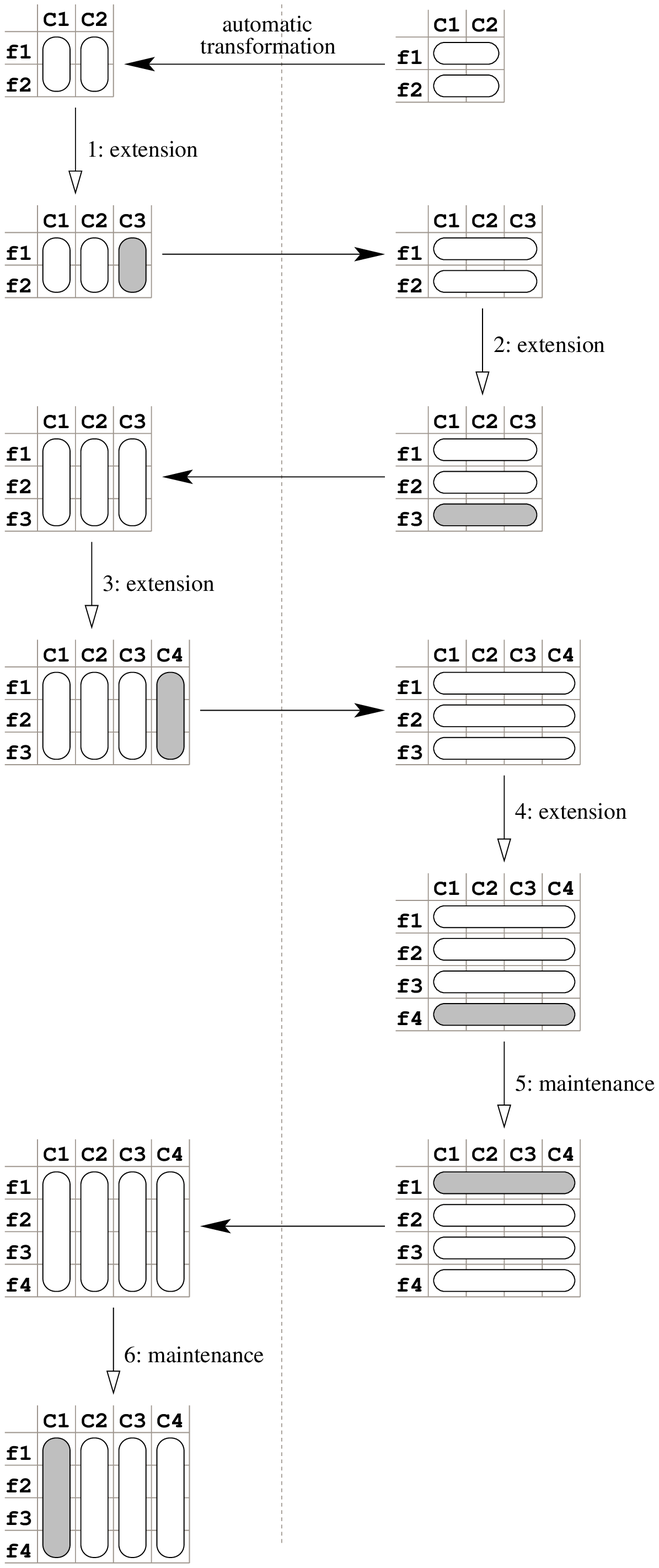}
\label{fig-ideal}
}
\caption{Progress of the evolution scenario in different cases.}
\label{fig:scenario}
\end{figure}

We now assume that we have chosen a particular solution to
extend any axis by providing a new module (for instance, one
of  the solutions cited in~\cite{Zenger-Odersky2005}) and we
examine what happens with the scenario of Fig~\ref{fig-scenario-definition}.
The progress of this scenario is illustrated by
Figure~\ref{fig-current-solutions} and is detailed
below. Grey zones in the figure represent the code which has
been added or modified.

%\begin{figure}[p]
%\begin{center}
%\includegraphics[scale=0.6]{figures/scenario_current2.eps}
%\end{center}
%\caption{scenario for some solutions to the extension problem}
%\small
%Grey zones represent new or modified code.

%\label{fig-current-solutions}
%\end{figure}

\paragraph{Two first extensions (evolutions 1 \& 2).}%%%%%%%%%%%%%%%%%%%%%%%%%%%%%%%%%%%
 After the first two evolutions, we are in one of the
following situations, depending on the initial program:\\

~\hfill
\includegraphics[scale=0.8]{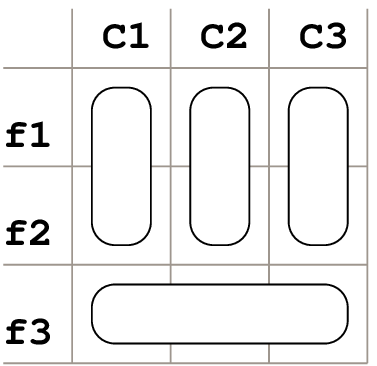} \hfill
\includegraphics[scale=0.8]{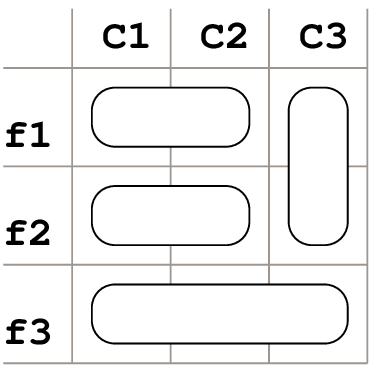}\hfill~\\

In the left-hand side diagram, the extension of the data
type with \texttt{C3} is natural, and adding the function
\texttt{f3} can be done with the chosen specific language feature
(in this case, the ``module'' for \texttt{f3} has a different nature
from the three other modules of the application).
%
% \fbox{REMI: la matrice a t elle du sens avec les extensions language?}
%

In the right-hand side diagram, we have extended
\texttt{f1} and \texttt{f2} with the chosen specific modular
feature to take \texttt{C3} into account and then we add \texttt{f3}.
If we want the extension for \texttt{f3} to
be fully modular, we have to define \texttt{f3} on
\texttt{C1}, \texttt{C2} and \texttt{C3} in a single
module.
(Another solution would have been to make a module
with \texttt{f3} defined on \texttt{C1} and \texttt{C2} and
to complete the module of \texttt{C3}, but we do not consider this is modular).
Even if the modules for \texttt{f1}, \texttt{f2} and \texttt{f3} are of the same nature, they do not cover the same subset of constructors.

This means that one
cannot fully rely on \texttt{f1} or \texttt{f2} as patterns
to write \texttt{f3} (\emph{problem 1: loss of regularity for extensions}).

\paragraph{Two following extensions (evolutions 3 \& 4).}%%%%%%%%%%%%%%%%%%%%%%%%%%
 Now, let us take two more extensions into account.\\

~\hfill\includegraphics[scale=0.8]{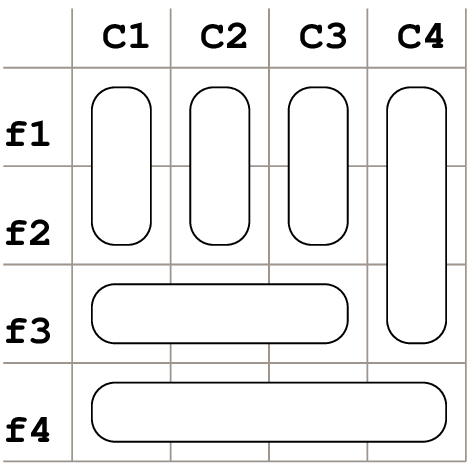} \hfill
\includegraphics[scale=0.8]{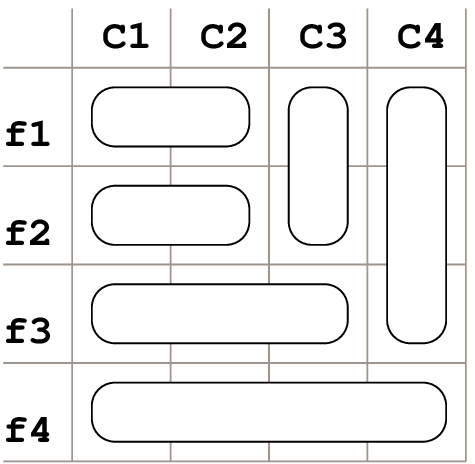}\hfill~\\

In the left-hand side diagram, \texttt{C4} is added
naturally as a module, but we see that the corresponding
module does not cover the same functions as the modules for
\texttt{C1}, \texttt{C2} and \texttt{C4} (this boils downs
to the problem~1). Then \texttt{f4} is added with the same
technique as \texttt{f3} but, again, the module for
\texttt{f4} does not cover the same cases as the module for
\texttt{f3}.

We meet the same problems in the case of the right-hand side diagram.

We can observe that the regular architecture of the initial
programs rapidly becomes disordered with incremental
extensions. This will reveal to be bad at maintenance time.

\paragraph{Maintenance time (evolutions 5 \& 6).}%%%%%%%%%%%%%%%%%%%%%
Now we have to modify \texttt{f1} (to correct an error or to cope with
a change in its specification).  In the (initially) data-centered architecture
(left-hand side), %we have nothing good to hope for as 
the code for \texttt{f1} is already spread over several modules in the original
program.
In the (initially) operation-centered architecture, the code
is finally also spread over several modules. This means
that we have lost the benefit of the initial modularity:
the maintenance is no more modular (\emph{problem 2: loss of the initial modularity properties}).

This is the same for the maintenance of \texttt{C1}: in the
data-centered architecture, the code has become spread
over several modules. 
Moreover, the number of modules on which the code dealing with a constructor is spread is different for \texttt{C1} and \texttt{C3}.

%Furthermore, as two modules of the same nature (data- or
%function-centered) cover different cases, the maintenance of a
%function  has not the same properties with respect to
%modularity as the maintenance of an other function.
%
%For instance, in the operation-centered architecture (right
%hand side diagram), modifying \texttt{f3} requires to modify
%(at most) 2 modules whereas modifying \texttt{f2} requires
%to modify 3 modules. The same is true for constructors. The maintenance task has lost the
%regularity it had in the initial programs (which is a consequence of the problem 2).
%
%\fbox{REMI: quelle difference avec le problem 2?}

The example of this section shows that the technical
solutions for modular extensibility are not sufficient for
modular maintainability.

\section{Views and Transformations}%%%%%%%%%%%%%%%%%%%%%%%%%%%%%%%%%%%%%%
\label{sec-transformation}
Multiple views~\cite{Black2004} aim at solving the problem
described in the previous section. We now recast this in our setting.

\subsection{Programs and Views}

We call \emph{views} of a program two or more textual
representations of programs that have the same behavior (they
are semantically equivalent in a given calculus).
For instance, $\pdata$ and $\pfun$ are two views of the same program
(this is justified later in the paper).

%For a given relation of equivalence, we can see a program as a set
%of source texts of the considered language which are equivalent (\ie a
%class of equivalence).
%
%For a given program, or for a given class of equivalence, we
%call \emph{views} of the program the elements of the class,
%\emph{i.e.} textual representations of that program.
%Of course, in general relations of equivalence are undecidable (\eg
%does two programs define the same function), but we will focus on
%views for which the equivalence is decidable (otherwise, we could not
%prove that our transformation is correct).

%Following these definitions, $\pdata$ and $\pfun$ are
%two views of the same program (this is justified later in the paper).

\subsection{A Solution to the Modular Maintenance Problem}%%%%%%%%%%%%%%%%%%%%%%%%

\label{sec-solution}
 We illustrate the use of views to solve the problem of
 modularity in evolutions by building an automated
 transformation of $\pdata$ into $\pfun$ and its reverse
 transformation.

%To illustrate this, we have developed an automatic
%transformation between $\pdata$ into $\pfun$, along with the reverse
%transformation.
%
%
%tool such that if we have a data-centered view of a program,
%the tool can compute the operation-centered view of the same
%program, and vice-versa. In particular, in our example of
%section~\ref{sec-primary-decomposition}, the tool is
%able to transform $\pdata$ into $\pfun$ and the other way around.
%
With such a tool, the programmer can choose the view in
which the evolution he has to implement is modular.
%
%%%%%%%%%
%
Fig.~\ref{fig-ideal} illustrates the scenario
given in section~\ref{sec-scenario} with this approach.
For instance, when the programmer wants to add a new
constructor (evolution 1), the program is presented in
the data-centered view;
when the programmer wants to add a new function (evolution 2),
the program is presented in the operation-centered
view.
Since none of these evolutions has to be made transversely to the considered
axis of decomposition, the views of interest always keep a regular architecture.
This approach enables to solve the problems 1
and 2 described in section~\ref{sec-problem}.

This approach also has the following advantages compared to solutions to the expression problem:
\begin{itemize}

\item It does not rely on a particular programming language. As soon as two alternative programming structures can be expressed in a language, the corresponding transformation can be defined. 

\item The programmer who implements the evolution
  does not have to learn a new language or possibly complex language
  features. Of course, he has to cope with several views.

\item The programmer does not have to learn a new type system. In particular, if the programming language is strongly typed, the different views are also strongly typed and the types they introduce are closely related. In this case, typing issues boil down to verify once for all that the program transformations do not break typing.
%\emph{Can we say something about typing in the two views, do we have the same types? Some types might be mode general in one of the two views? We can aggregate the properties of the two views?}

\item The approach is not limited to the data-centered view \textit{versus} the function-centered view. It is not even limited to two views.

\end{itemize}

Of course, this approach is not free from disadvantages:\begin{itemize}

\item The programmer who implements the evolution has to cope with several views.

\item The transformation has to be implemented, which requires some work from a ``transformation designer'' and a supporting tool.

\end{itemize}

%We do not explore in this paper the different possibilities to build
%such a tool.
%
%Rather, we report our experience in building the $\pfun \longleftrightarrow \pdata$ transformation by relying on (and adapting) a refactoring tool.
%

\subsection{Refactoring tools to navigate between views}%%%%%%%%%%%%%%%%%%%%%%%%
\label{refactoring-to-navigate}

Developing a program transformation from scratch is not easy. 
Refactoring tools provide a simple, high-level way to transform
programs and are available for several kinds of languages. 
For this reason, we have chosen to explore the use of a refactoring
tool to build our example of transformation.

\emph{``Refactoring is the process of changing a software system in such a way that it does not alter the external behavior of the code yet improves its internal structure''}~\cite{Fowler1999}

Given our definition of programs and views, a code refactoring changes
the views 
of the considered program (a refactoring tool enables  
to pass
from one view to another). 
Griswold~\cite{Griswold1991} shows 
that
refactoring tools can be used to change a function-centered
architecture into a data-centered architecture. In
particular, he exemplifies his technique with
the program
Parnas~\cite{Parnas1972} had employed to illustrate
different
architectures for a same program, each
with different advantages and disadvantages.
Griswold uses refactoring tools to improve the structure of code. 
We adopt a more dynamical point of view,
where an improvement is not absolute, but driven by a
temporary need:  once the driving evolution is implemented, we
may need to revert to the initial architecture.

Some refactoring tools are provided with most popular IDEs, for several mainstream languages (Java in Eclipse~\cite{Eclipse-unsound} and NetBeans~\cite{netbeans}, C++, C\#, VB in Visual Studio~\cite{VisualAssistX}).
Refactoring tools have also found an interest in the academic
community, and some tools which are based on sound
foundations have been proposed, for instance for Haskell and Erlang~\cite{HaRe-Wrangler2008}, C~\cite{CScout2010}, Smalltalk~\cite{Smalltalk-refactoring-1997} or Lisp~\cite{Griswold1993}.
Finally, some refactoring tools have been specifically designed to support views in an object-oriented context~\cite{Black2004,Shonle2007} (see section~\ref{related-work}).

%If each of these operations is correct, that is if the input
%and the result are functionally equivalent, applying
%successively elementary refactoring operations is also
%correct since equivalence is transitive. For this
%reason, refactoring tools allow to navigate into classes of
%equivalence of programs. In the general case, the whole
%class cannot be accessed by this means, but it is not
%important in our case.

It is important to note that some refactoring tools are not
sound.
In particular, with the Eclipse refactoring tool for Java, after
refactoring operations have been applied, the user has to fix broken
code himself~\cite{Eclipse-unsound,Eclipse-unsound2009}.
For this reason, we focus in this paper on one of the refactorers which provides operations whose principles 
have been proved correct:
%:\footnote{Although the principles has been
%  proved correct, there are some bugs left in the interface, and
%  possibly in the implementation.} 
the Haskell Refactorer
(HaRe)~\cite{HaRe-web, HaRe-formalisation, Sultana-Thompson2008}.
The formalism used in~\cite{HaRe-formalisation} is a
$\lambda$-calculus with \emph{let-rec} with a mixed
call-by-name/call-by-need strategy. 
A relation of equivalence based on the reduction rules of
that calculus expresses the behavior preservation which is
used to prove the correctness of the operations.

Before applying a refactoring operation, the Haskell Refactorer 
checks that the conditions to ensure its correctness are verified. For instance, it checks that a
renaming does not introduce a name clash. When these conditions
are not verified, the refactorer does not apply the changes
and explains why.

%\emph{``If the refactoring causes problems in other methods, these are ignored and you must fix them yourself after the refactoring.''}~\cite{Eclipse-unsound}

%\emph{``Note that some modifications you make to the method, such as adding a parameter or changing a return type, may cause the refactored code to contain compiler errors because Eclipse doesn't know what to enter for those new parameters.''}~\cite{Eclipse-unsound2009}

%\begin{boxedminipage}{12cm}

%HaRe has the good property that it will check whether a
%transformation is correct in the considered context before
%applying it~\cite{HaRe-formalisation}.
%
%For this reason, we do not have to provide the proof of
%correctness for each elementary step of our transformation
%if it is handled by HaRe.
%\end{boxedminipage}

\section{Implementation of an Architecture Transformer with a Refactorer}%%%%%%%%%%%%%%%%%%
\label{sec-refactoring-haskell}
In this section, we show how $\pfun$ can be transformed into $\pdata$ by using the Haskell Refactorer, and the other way around. 
We also show how the chain of refactoring operations can be automated.

%
%\emph{This transformation could have been defined as a global
%correspondence between two patterns of programs,
%
%but we prefer to define it by a sequence of elementary
%transformations, following the divide and conquer strategy.
%
%
%
%Using a sequence of refactoring steps instead of a global
%transformation has the advantage that each elementary step
%is easy to understand, to prove correct (if not already
%done), to maintain and to reuse.}
%
% The same is true for the whole sequence of refactorings which architecture is very simple.

%As a first experiment, our implementation is not very generic and can be improved, see the discussion in section~\ref{discussion-generalite}.

\subsection{Decomposing the transformation into refactoring operations}%%%%%%%%%%%%%%%%%%%%%%%%%%%%%%%%%%%
\label{transformation-steps}
We describe in this section the steps to transform $\pfun$ into $\pdata$ with the Haskell Refactorer. 
As already said, each of these operations checks that the conditions that
make the refactoring correct are verified. 
%
%For instance, potential name clashes are detected.

In the following, we consider only the \texttt{eval} function. The chain of
operations is the same on the \texttt{toString} function. Here is
the code which is considered in $\pfun$:

\begin{lstlisting}
-- from EvalMod
eval (Const i)     = i
eval (Add (e1,e2)) = eval e1 + eval e2
\end{lstlisting}
\begin{lstlisting}
--from Client
r2 = print (show (eval e1))
r4 = print (show (eval e2))
\end{lstlisting}

We now present the transformation steps. All the fragments of code we show in this section are the result of the use of the refactorer (except for the comments introduced by \texttt{--} and some empty lines which are skipped).

\begin{enumerate}

\item %%%%%%%%%%%%%%%
We introduce new local definitions for the code of each constructor case (Haskell Refactorer's  \emph{Introduce New Definition}  operation).

\begin{lstlisting}
eval (Const i)     = evalConst
  where
    evalConst = i
eval (Add (e1,e2)) = evalAdd
  where
    evalAdd = (eval e1) + (eval e2)
\end{lstlisting}

\item %%%%%%%%%%%%%
 In each of these new definitions, depending on the constructor concerned
 (\texttt{Const} or \texttt{Add}), we generalize either the arguments
 of the constructor, or the recursive calls of \texttt{eval} on these
 arguments (\emph{Generalise def}).

\begin{lstlisting}
eval (Const i)     = evalConst i
  where
    evalConst x = x
eval (Add (e1,e2)) = evalAdd (eval e1) (eval e2)
  where
    evalAdd x y = (x) + (y)
\end{lstlisting}

\item %%%%%%%%%%%%%%%%%
We lift the new local definitions to make them global (\emph{Lift Definition to Top Level}).
\begin{lstlisting}
eval (Const i)     = evalConst i
  
eval (Add (e1,e2)) = evalAdd (eval e1) (eval e2)

evalAdd x y = (x) + (y)
  
evalConst x = x
\end{lstlisting}

\item %%%%%%%%%%%
In the definition of \texttt{eval}, we generalize \texttt{evalConst} and \texttt{evalAdd} (\emph{Generalise def}).

\begin{lstlisting}
eval a c (Const i)     = c i
  
eval a c (Add (e1,e2)) = a (((eval a) c) e1) (((eval a) c) e2)
\end{lstlisting}

By applying that operation, all the existing references
to \texttt{eval}, in particular in the module Client, are transformed
to take the new parameters into account. In the body of \texttt{eval},
references to \texttt{eval} are replaced by \texttt{((eval f\_Add)
f\_Const)}.
In the module \texttt{Client}, references to \texttt{eval} are
replaced by \texttt{eval eval\_gen\_1 eval\_gen} where \texttt{eval\_gen}
is defined by \texttt{evalConst} in \texttt{EvalMod}
and \texttt{eval\_gen\_1} is defined by \texttt{evalAdd}.

\begin{lstlisting}
-- from EvalMod module
eval_gen_1 = evalAdd

eval_gen = evalConst
\end{lstlisting}
\begin{lstlisting}
-- from Client module
r2 = print (show (eval eval_gen_1 eval_gen e1))
r4 = print (show (eval eval_gen_1 eval_gen e2))
\end{lstlisting}

\item %%%%%%%%%%

We rename \texttt{eval} into \texttt{fold1} (\emph{Rename}).

\begin{lstlisting}
-- from EvalMod module
fold1 a c (Const i)     = c i
  
fold1 a c (Add (e1,e2)) = a (((fold1 a) c) e1) (((fold1 a) c) e2)
\end{lstlisting}
\begin{lstlisting}
-- from Client module
r2 = print (show (fold1 eval_gen_1 eval_gen e1))
r4 = print (show (fold1 eval_gen_1 eval_gen e2))
\end{lstlisting}

\item %%%%%%%%%%%%%
We introduce a new definition named \texttt{eval} for the expression \texttt{fold1 eval\_gen\_1 eval\_gen e1} in the module \texttt{Client}, we lift it at the top level, and abstract it over \texttt{e1}.
\begin{lstlisting}
r2 = print (show (eval e1))

eval x = fold1 eval_gen_1 eval_gen x

r4 = print (show (fold1 eval_gen_1 eval_gen e2))
\end{lstlisting}

%\emph{attention, c'est un peu contre-intuitif, uniquement dans le but de faciliter l'ingenierie de l'automatisation. Ici on avait envie de generaliser tout de suite e1 au lieu de lifter. Pour ce faire, il faudrait introduire une operation qui fasse un generalise-indent sans l'argument qui indique dans quelle top-level-declaration la generalisation doit avoir lieu.}

\item %%%%%%%%%%%%%

We fold the definition of \texttt{eval} to make it appear in \texttt{r4} (\emph{Fold Definition}).

\begin{lstlisting}
r2 = print (show (eval e1))

eval x = fold1 eval_gen_1 eval_gen x

r4 = print (show (eval e2))
\end{lstlisting}

\item %%%%%%%%%%%

We unfold the occurrences of \texttt{eval\_gen}
and \texttt{eval\_gen\_1} (\emph{Unfold def}) and we remove their
definitions (\emph{Remove def}).

\begin{lstlisting}
eval x = fold1 evalAdd evalConst x
\end{lstlisting}

\item %%%%%%%%%%%
We move the definitions of \texttt{evalConst} and \texttt{evalAdd} from \texttt{EvalMod} to \texttt{ConstMod} and \texttt{AddMod} and rename them into \texttt{eval}.

\begin{lstlisting}
r2 = print (show (Client.eval e1))

eval x = fold1 AddMod.eval ConstMod.eval x
  
r4 = print (show (Client.eval e2))
\end{lstlisting}

\item %%%%%%%%%%%%%
We move the definition of \texttt{fold1} into the \texttt{Expr} module. The module \texttt{EvalMod} is now empty.

\item %%%%%%%%%%%%%
We remove useless imports of module \texttt{EvalMod} in the module \texttt{Client} (\emph{Clean imports}).

\end{enumerate}

In practice, after this sequence of refactorings,
\texttt{fold1} and the \texttt{fold2} we get 
from the transformation of \texttt{toString}  are $\alpha$-equivalent. 
One of them may be deleted to find the exact $\pdata$
described in section~\ref{sec-primary-decomposition} (this
seems not to be supported by the Haskell Refactorer at the
moment).

The layout is also not exactly the same as expected 
%(the declarations are not in the same order and 
(\eg there are additional pairs of parenthesis).

%Note that this transformation, supported by the Haskell Refactorer, shows that $\pfun$ and $\pdata$ are semantically equivalent.

\paragraph{Soundness.} 
If we use behavior preserving refactoring operations,
then the chain of refactoring operations is also behavior preserving (and $\pfun$ is equivalent to $\pdata$).
Some Haskell Refactorer's operations's principles have been shown to be correct~\cite{Sultana-Thompson2008, HaRe-formalisation}.
However, there are some bugs left in the implementation, and not all
the available operations have been proved correct, including some of
the ones we are using\footnote{This problem is
  different from the problem in Eclipse for which the answer is :
\emph{``If the refactoring causes problems in other methods, these are
  ignored and you must fix them yourself after the
  refactoring.''}~\cite{Eclipse-unsound} and
\emph{``Note that some modifications you make to the method, such as
  adding a parameter or changing a return type, may cause the
  refactored code to contain compiler errors because Eclipse doesn't
  know what to enter for those new
  parameters.''}~\cite{Eclipse-unsound2009}.  }

% liste des operations utilisees (fun to data)
% - introduce new def
% - generalise definition
% - lift def (to top level)
% - rename
% - move def between modules
% - unfold
% - remove (unused) definition 

% liste des operations utilisees (data to fun)
% - rename 
% - duplicate into comment (ajoute)
% - unfold
% - generative fold (Burstall Darlington)
% - remove comment (ajoute)
% - simplify case (ajoute)
% - remove (unused) definition
% - case to equation (ajoute)
% - clean imports + rm from exports ou move def between modules, avec quand meme un clean imports
% - optional : demote def

% liste des operations prouvees correctes parmi celles utilisees
% - generalise
% - move def between module
% - *** trouver une exemple de remove duplicate def ***
% - demote def
% - introduce new def

% liste des operations pas prouvees
% - lift
% - rename (mais dans thèse de Li ?)
% - unfold 
% - remove unused
% - toutes celles que j'ai ajoute
% - generative fold (deja prouve par ailleurs?)
% - clean imports, remove from exports

% question : si c'est standard ou prouve dans le lambda-calsul, peut-on considérer que c'est rpouvé dans le cadre de haskell? Pour le clean imports, le système est spécifique à Haskell, il faut certainement prouver des choses spécifiques. Par contr, le unfold doit etre ultra classique (beta reduction?)  
 
%\emph{Conclusion? les outils vont murir? la correction du refactorer est hors du sujet de cet article? On ne sait pas si les outils vont murir car ils sont finalement peu utilises (Murphy-Hill/Black : Why Don't People Use Refactoring Tools)? S'agit-il d'un inconvenient a rappeler dans le bilan?}

\paragraph{Reverse transformation.}%%%%%%%%%%%%%%%%%%%%%%%%%%%%%%%%%%
A simple approach to build the reverse transformation would be to
use the inverse of each operation used in the
$\pfun \rightarrow \pdata$ transformation in the reverse order (since $(f \circ g)^{-1} =
g^{-1} \circ f^{-1}$). 
However, the Haskell Refactorer does not provide an inverse
for each operation, so our reverse
transformation cannot be automatically derived from the first
transformation.

We do not detail here the reverse transformation (the script is given in Figs~\ref{fig-reverse-script} and~\ref{fig-reverse-script-cont} ).
The key steps are to unfold the instances of \texttt{ConstMod.eval},
\texttt{ConstMod.toString}, \texttt{AddMod.eval},
\texttt{AddMod.toString} and to transform \texttt{eval} and
\texttt{toString}, which are defined by calls to \texttt{fold\_1}
and \texttt{fold\_2}, into plain recursive function definitions.
This particular step is done by using the \emph{Generative
Fold} operation of the Haskell Refactorer (\emph{folding} in~\cite{Burstall1977}, see~\cite{Brown2008}).
Note that to use the generative fold, a preliminary step has to be
done by hand: a function definition must be duplicated into a comment.
We have introduced in the Haskell Refactorer a feature to support this
duplication into comments in order to make the whole chain supported
by the tool and to be able to automate it.

We also have had to add some features to the refactorer to simplify the
code we get after unfolding some functions defined by
equations with patterns in order to reach the code of $\pdata$ (see appendix~\ref{app-added-operations} for the list of  operations we have implemented into HaRe).

\subsection{Automation of the process.}%%%%%%%%%%%%%%%%%%%%%%%%%%%%%%%%%%%%%
%%%%%%%%%%%%%%%%%%%%%%%%%%%%%%%%%%%%%%%%%%%%%%%%%%%%%%%%%%%%%%%%%%%%%%%%%%%%
\lstset{basicstyle=\small, frame=trBL, frameround=tfff, language=lisp, basicstyle=\ttfamily}
%We have automated the transformation by using successive
%calls to the refactorer operations. 

An engineering effort has been necessary to automate the
transformation since the API of the Haskell Refactorer does not match
our needs.
In particular, HaRe is designed to be used interactively with text editors as Emacs and the
parameters of the operations are cursor positions (line and column numbers) in
source files.
For this reason we have developed functions to locate
sub-expressions of interest in files before calling HaRe
operations with the computed parameters. 
This allows us to provide new interfaces for HaRe operations
(at least for those we needed) that do not rely on the
particular layout in the source files (see appendix~\ref{app-added-interfaces} for the list of interfaces to HaRe operations we have implemented).
Like original HaRe operations, our interfaces to HaRe operations are available as Emacs-Lisp
functions in addition to interactive menu entries. 
This allows to invoke them in Emacs-Lisp programs.

Our transformations can thus be expressed by Emacs-Lisp programs. Since these programs are reduced to sequences of operations with side-effects, we call them \emph{scripts}.
Fig.~\ref{fig-script} shows the script of the transformation $\pfun \rightarrow \pdata$ (we assume the definitions of Fig.~\ref{fig-script-definitions} are evaluated first).

\begin{figure}[htp]
\relsize{-2}
\begin{lstlisting}[firstline=1,lastline=17]
(defvar f1        "eval"           )
(defvar f2        "toString"       )
(defvar c1        "Const"          )
(defvar c2        "Add"            )
(defvar f1mod     "EvalMod"        )
(defvar f2mod     "ToStringMod"    )
(defvar f1c1      (concat f1 c1)   )
(defvar f1c2      (concat f1 c2)   )
(defvar f2c1      (concat f2 c1)   )
(defvar f2c2      (concat f2 c2)   )
(defvar c1mod     (concat c1 "Mod"))
(defvar c2mod     (concat c2 "Mod"))
(defvar f1reducer "fold1"          )
(defvar f2reducer "fold2"          )
(defvar dummyc1   "c"              )
(defvar dummyc2   "a"              )
(defvar clientmod "Client"         )
\end{lstlisting}
\caption{$\pfun \rightarrow \pdata$ script preliminary definitions}
\label{fig-script-definitions}
\normalsize
\end{figure}

\begin{figure}[htp]
\relsize{-2}
\begin{lstlisting}[firstline=19]
;; 1 - introduce new definitions
(haskell-refac-exhibitFunction f1 c1 f1c1 f1mod)
(haskell-refac-exhibitFunction f1 c2 f1c2 f1mod)
(haskell-refac-exhibitFunction f2 c1 f2c1 f2mod)
(haskell-refac-exhibitFunction f2 c2 f2c2 f2mod) 

;; 2 - abstract some arguments in these definitions
(haskell-refac-generalise f1 c1 f1c1 f1mod "0" "x" "Curried" "OtherType")
(haskell-refac-generalise f2 c1 f2c1 f2mod "0" "x" "Curried" "OtherType")
(haskell-refac-generalise f2 c2 f2c2 f2mod "1" "y" "UnCurried" "RecType")
(haskell-refac-generalise f2 c2 f2c2 f2mod "0" "x" "UnCurried" "RecType")
(haskell-refac-generalise f1 c2 f1c2 f1mod "1" "y" "UnCurried" "RecType")
(haskell-refac-generalise f1 c2 f1c2 f1mod "0" "x" "UnCurried" "RecType") 

;; 3 - lift the new functions to the top-level
(haskell-refac-makeGlobalOfLocalIn f1 f1c1 f1mod)
(haskell-refac-makeGlobalOfLocalIn f1 f1c2 f1mod)
(haskell-refac-makeGlobalOfLocalIn f2 f2c1 f2mod)
(haskell-refac-makeGlobalOfLocalIn f2 f2c2 f2mod) 

;; 4 - abstract the functions of interest from the introduced functions 
(haskell-refac-generaliseIdent f1 f1mod f1c1 dummyc1)
(haskell-refac-generaliseIdent f1 f1mod f1c2 dummyc2)
(haskell-refac-generaliseIdent f2 f2mod f2c1 dummyc1)
(haskell-refac-generaliseIdent f2 f2mod f2c2 dummyc2) 

;; 5 - rename the functions of interest (they have become traversal functions)
(haskell-refac-renameToplevel f1 f1mod f1reducer)
(haskell-refac-renameToplevel f2 f2mod f2reducer) 

;; 6 - reconstruct the functions of interest as calls to the traversal functions
;; with appropriate arguments
(haskell-refac-newDefFunApp f1reducer "3" f1 clientmod)
(haskell-refac-newDefFunApp f2reducer "3" f2 clientmod)
(haskell-refac-makeGlobalOfLocal f1 clientmod)
(haskell-refac-makeGlobalOfLocal f2 clientmod) 
(haskell-refac-generaliseIdent f1 clientmod "e1" "x")
(haskell-refac-generaliseIdent f2 clientmod "e1" "x") 

;; 7 - propagate the new definitions
(haskell-refac-foldToplevelDefinition f1 clientmod)
(haskell-refac-foldToplevelDefinition f2 clientmod)

;; 8 - unfold dummy definitions and remove them 
(haskell-refac-unfoldInstanceIn (concat f1 "_gen") f1 clientmod)
(haskell-refac-unfoldInstanceIn (concat f1 "_gen_1") f1 clientmod)
(haskell-refac-unfoldInstanceIn (concat f2  "_gen") f2 clientmod)
(haskell-refac-unfoldInstanceIn (concat f2  "_gen_1") f2 clientmod) 

(haskell-refac-removeDefCmd (concat f1 "_gen") f1mod)
(haskell-refac-removeDefCmd (concat f1 "_gen_1") f1mod)
(haskell-refac-removeDefCmd (concat f2 "_gen") f2mod)
(haskell-refac-removeDefCmd (concat f2 "_gen_1") f2mod)

;; 9 - move business functions to the appropriate modules
(haskell-refac-moveDefBetweenModules f1c1 f1mod c1mod)
(haskell-refac-moveDefBetweenModules f1c2 f1mod c2mod)
(haskell-refac-moveDefBetweenModules f2c1 f2mod c1mod)
(haskell-refac-moveDefBetweenModules f2c2 f2mod c2mod)

;; rename the business functions to make them share the same name
(haskell-refac-renameToplevel f1c1 c1mod f1)
(haskell-refac-renameToplevel f1c2 c2mod f1)
(haskell-refac-renameToplevel f2c1 c1mod f2)
(haskell-refac-renameToplevel f2c2 c2mod f2) 

;; 10 - move the functions into the relevant modules
(haskell-refac-moveDefBetweenModules f1reducer f1mod "Expr")
(haskell-refac-moveDefBetweenModules f2reducer f2mod "Expr")

;; 11 - clean imports
(haskell-refac-cleanImportsCmd clientmod)
\end{lstlisting}
\caption{$\pfun \rightarrow \pdata$  transformation script}
\label{fig-script}
\normalsize
\end{figure}

%The interface we have added to HaRe can be used to automate other
%refactoring-based transformations.
%
%However, all the needs to automate a refactoring-base transformation
%is not necessarily covered by our current interface (\emph{décrire les operations fournies, comparer aux operations necessaires pour KWIC}).
%
%The same is true for the refactoring operations we have had to implemented.

%The script of the reverse transformation is given in figure~\ref{fig-reverse-script}.

%\emph{The source code of our interfaces, operations and examples of scripts are available
%at \thisarticlepage ~.}

\begin{figure}[!htp]
\relsize{-2}
\begin{lstlisting}[firstline=19, lastline=99]
;; reverse 10 -
(haskell-refac-moveDefBetweenModules f1reducer "Expr" f1mod)
(haskell-refac-moveDefBetweenModules f2reducer "Expr" f2mod)

;; reverse 9 - use specific names for business functions 
(haskell-refac-renameToplevel f1 c1mod f1c1)
(haskell-refac-renameToplevel f1 c2mod f1c2)
(haskell-refac-renameToplevel f2 c1mod f2c1)
(haskell-refac-renameToplevel f2 c2mod f2c2)

;; reverse 9 - move business functions to the original modules 
(haskell-refac-moveDefBetweenModules f1c1 c1mod f1mod)
(haskell-refac-moveDefBetweenModules f1c2 c2mod f1mod)
(haskell-refac-moveDefBetweenModules f2c1 c1mod f2mod)
(haskell-refac-moveDefBetweenModules f2c2 c2mod f2mod)

;; move f1 f2 to f1mod f2 mod
(haskell-refac-moveDefBetweenModules f1 clientmod f1mod)
(haskell-refac-moveDefBetweenModules f2 clientmod f2mod)

(haskell-refac-cleanImportsCmd clientmod) 

;; reverse 4/5/6/7 - transform a call to fold into a recursive function

;; prepare the generative fold operations :
;; the equations for functions of interest are saved into  comments
(haskell-refac-duplicateIntoComment f1 f1mod)
(haskell-refac-duplicateIntoComment f2 f2mod)

;; unfold the use of the traversal function
;; in the definition of functions of interest
(haskell-refac-unfoldInstanceIn f1reducer f1 f1mod)
(haskell-refac-unfoldInstanceIn f2reducer f2 f2mod)

;; unfolding has produced case expressions that have
;; to be simplified
(haskell-refac-simplifyCasePattern f1 f1mod)
(haskell-refac-unfoldInstanceIn dummyc2 f1 f1mod)
(haskell-refac-unfoldInstanceIn dummyc2 f1 f1mod)
(haskell-refac-unfoldInstanceIn dummyc2 f1 f1mod)
(haskell-refac-removeLocalDef dummyc2 f1 f1mod)

(haskell-refac-simplifyCasePattern f1 f1mod)
(haskell-refac-unfoldInstanceIn dummyc1 f1 f1mod)
(haskell-refac-unfoldInstanceIn dummyc1 f1 f1mod)
(haskell-refac-unfoldInstanceIn dummyc1 f1 f1mod)
(haskell-refac-removeLocalDef dummyc1 f1 f1mod)

(haskell-refac-simplifyCasePattern f2 f2mod)
(haskell-refac-unfoldInstanceIn dummyc2 f2 f2mod)
(haskell-refac-unfoldInstanceIn dummyc2 f2 f2mod)
(haskell-refac-unfoldInstanceIn dummyc2 f2 f2mod)
(haskell-refac-removeLocalDef dummyc2 f2 f2mod)

(haskell-refac-simplifyCasePattern f2 f2mod)
(haskell-refac-unfoldInstanceIn dummyc1 f2 f2mod)
(haskell-refac-unfoldInstanceIn dummyc1 f2 f2mod)
(haskell-refac-unfoldInstanceIn dummyc1 f2 f2mod)
(haskell-refac-removeLocalDef dummyc1 f2 f2mod)

;; the case expressions introduced by the unfolding 
;; have to be transformed into equations
(haskell-refac-caseToEq f1 f1mod)
(haskell-refac-caseToEq f2 f2mod)

;; fold the use of the traversal function in the body of the
;; functions of interest in order to get a recursive definition 
;; (without a call to the traversal function)
(haskell-refac-generativeFold f1reducer "3" f1mod)
(haskell-refac-generativeFold f1reducer "3" f1mod)
(haskell-refac-generativeFold f2reducer "3" f2mod)
(haskell-refac-generativeFold f2reducer "3" f2mod)

;; comments introduced for the generative fold can be deleted
(haskell-refac-rmCommentBefore f1 f1mod)
(haskell-refac-rmCommentBefore f2 f2mod)

;; the traversal functions can be deleted
(haskell-refac-removeDefCmd f1reducer f1mod)
(haskell-refac-removeDefCmd f2reducer f2mod)
;; end of the transformation of the call to fold into a recursive function.
\end{lstlisting}
\caption{$\pdata \rightarrow \pfun$  transformation script (first part)}
\label{fig-reverse-script}
\normalsize
\end{figure}

\begin{figure}[!htp]
\relsize{-2}
\begin{lstlisting}[firstline=101]
;; reverse 1/2/3 - replace calls to the business functions by their bodies (unfold)
;; and delete the business functions.
(haskell-refac-unfoldInstanceIn f1c1 f1 f1mod)
(haskell-refac-unfoldInstanceIn f1c2 f1 f1mod)
(haskell-refac-unfoldInstanceIn f2c1 f2 f2mod)
(haskell-refac-unfoldInstanceIn f2c2 f2 f2mod)

(haskell-refac-removeDefCmd f1c1 f1mod) 
(haskell-refac-removeDefCmd f1c2 f1mod)
(haskell-refac-removeDefCmd f2c1 f2mod)
(haskell-refac-removeDefCmd f2c2 f2mod)


\end{lstlisting}
\caption{$\pdata \rightarrow \pfun$  transformation script (end)}
\label{fig-reverse-script-cont}
\normalsize
\end{figure}

\section{Usage and Future Work}%%%%%%%%%%%%%%%%%%%%%%%%%%%%%%%%%
This section discusses the possibility of using views with
the tool we demonstrated here and proposes future
works to improve this use.

%\subsection{Architectures equivalentes}
%-Exemples d'architectures equivalents

%-decidabilite d'equivalence

\label{sec-usage}

\paragraph{Development of the transformation.}

The first problem with our setting is that the
transformation has to be implemented. 
%In our example, the script is bigger than the initial program.  
We could ease this task by several means:
 \begin{itemize}

\item Compute automatically the transformation based on the
  hints of the designer, as done in~\cite{Black2004}
  and~\cite{Shonle2007} or provide a more high level
  transformation specific language.

\item Provide a list of examples to be used as transformation patterns.

\item Once a transformation is defined,
generate automatically the inverse transformation.

\item Record the sequence of commands used in an interactive refactorer (as Eclipse does~\cite{Eclipse-unsound2009}).

\end{itemize}

\paragraph{Maintenance of the transformation.}
 As the program evolves, the transformation may need to
 evolve too. We could study how an evolution impacts a
 transformation. Some simple examples seem to be directly
 tractable: for instance, if the evolution to be implemented
 is the renaming of a function, we can imagine that the
 refactoring tool that propagates the renaming could also
 propagate it into the transformation script.

\paragraph{Duration.}
The transformation of our example takes about 15 seconds (on
a 2.8 GHz Intel Pentium R processor) and 7 seconds for the
reverse transformation. This is longer than the compile time of the program 
(less than 1 sec.). 
%However, if we recast this in an industrial context, evolutions are not implemented everyday. 
We can suppose that the transformations can be
integrated into the build process that takes place once a
new version of the code is released, so that when a
programmer has to implement an evolution, all the designed
views are available. Of course, this supposes all the views
have been designed together with the initial program.

%We can note also that some work on the tools can reduce that
%computation length. In our example of implementation, a
%pre-condition checking is done with each elementary
%operation. By computing a minimal pre-condition for the
%whole chain, as proposed
%by~\cite{composition-of-refactorings2004}, we can reduce
%drastically that checking time.

\paragraph{Availability of a tool for a given language.} We have chosen to use an existing refactoring tool instead or rebuilding one. 
We have produced 615 LOC to implement missing refactoring
operations and 1160 LOC for the new interface. This has to
be compared to the size of a refactoring tool (11 KLOC for
Arcum~\cite{Shonle2007}).
So our approach depends on the tools available for the
language of interest.

%wc -l
%   70 RefacAddDeclIntoComment.hs
%  263 RefacSimplifyCase.hs
%  282 RefacUnsafeRenaming.hs
%   92 TransfoCommands.hs
%  319 TransfoOperations.hs
%  750 TransfoUtils.hs

%\paragraph{Tool Completeness.} L'outil ne sera jamais complet par rapport à tous les besoin imaginable (\cite{composition-of-refactorings2004}).

\paragraph{Failures}
From an operational point of view, as each operation of the
chain requires some pre-conditions to be satisfied, it may
occur that the user is informed that the transformation
cannot be achieved only after some operations have been
applied.
In order to inform the user as soon as possible (statically instead of
dynamically) that the transformation cannot be applied, we could compute
the pre-condition of the transformation, for instance by following the work of Kniesel
and Koch~\cite{composition-of-refactorings2004}.
This would imply to provide a formal description of the pre-conditions
and effects (post-conditions or Kniesel and Koch's backward-descriptions) of all the
operations used.

\section{Conclusion}%%%%%%%%%%%%%%%%%%%%%%%%%%%%%%%%%%%%%%%%%%%%%%%%%%%%%

\subsection{Contributions.} 

The contributions of this report are the following:
\begin{itemize}

\item We give an example of use of multiple views in a functional programming language. That example %is the classical example coming with 
comes from the expression problem. % on which we consider a transformation between two architectures.

\item We implement that transformation with an existing refactorer, the Haskell Refactorer, and we automate it. To be able to do this, we extend the interface of the refactorer and add a few operations in it.

\end{itemize}

%In this article, we have shown that modularity is related to extensions but also to maintenance.
%We have discussed the limitations of the solutions to the expression problem which are devoted to extension only.
%
%We have seen that solutions to the expression problem do not
%fit in the evolutivity problem.

%We have 
%identified a notion of views of programs in order
%to propose a solution to the evolutivity problem. 
%
%We have proposed an approach for modular maintenance (and for tackling the tyranny of the primary decomposition in general) which is based on program transformation techniques. 
%
%In particular, we have shown how to build an actual tool for modular maintenance.
%Our tool relies on a refactorer and it enables the programmer to write and modify code always in the ``right'' (modular) architecture of the application.

\subsection{Comparison to view tools}

Compared to some other multiple views implementations (in particular~\cite{Black2004}, \cite{Shonle2007}) and to the implementation of~\cite{Griswold1993}, our approach has the following pros and cons:

\begin{itemize}

\item $\oplus$ We rely on a previously existing refactorer. We had to implement few lines of code to adapt it to support views. The engineering effort is very small compared to building a dedicated tool. Moreover, the basic operations are already proved correct (yet not all of them, and not the concrete implementation).

\item $\ominus$ The  expression of the transformation is rather low level. It is imperative (as in~\cite{Griswold1993}) while in~\cite{Black2004} and~\cite{Shonle2007} it is declarative. Moreover, it is not automatically invertible.

\item $\not =$ We focus on dealing with fragments of business code
  while~\cite{Black2004} focuses on classes and method
  interactions and~\cite{Griswold1991} focuses on
  data-structure encapsulation (at least in the example from
  Parnas).

\end{itemize}

\subsection{Related Work}
\label{related-work}
Offering to the programmer (or designer or specifier) an appropriate view is not a new idea.
Here is a review of some relevant related work.

Before Wadler introduced the expression problem~\cite{expPb}, he had already proposed the notion of views~\cite{DBLP:conf/popl/Wadler87}.  His view feature makes it possible
to use pattern matching with different representations of a
data structure. At compile time different views of a data
structure are defined and their relationships are specified
by rewriting rules.  At run-time, a single data structure is
maintained and pattern matching on different representations
are translated to accessor functions that compute a (partial) view from another one.
This is closely related to our problem, but we do not focus on data structures but on code
structures and we extend and maintain programs before run-time. 

Griswold~\cite{Griswold1991, Griswold1993} shows
that elementary refactoring operations can be chained to
provide architecture transformations. However, as already
said in section~\ref{refactoring-to-navigate}, while Griswold's goal is to improve the
structure of code, our goal is to dynamically adopt a
structure which is convenient for a given task.

We share with Black and Jones~\cite{Black2004} the same
motivation and the idea that multiple views solves the tyranny of dominant decomposition
problem.
However, the techniques are rather different. In their
implementation, the programmer describes properties of the
fragments of code which are used to compute the views.
In Shonle \emph{et al.}~\cite{Shonle2007}, it is the
transformation which is described. It is described
declaratively by rewriting rules. Both works handle mainly
object oriented language concepts (classes, methods, field
accesses).

Functional programs are prone to be transformed. Numerous program
transformations have been proposed. Some comes in pairs, or are invertible.
For instance, Danvy have studied relationship between the continuation passing style and the direct style \cite{190866}.
In general a program transformation is not invertible. Forster et al. \cite{DBLP:journals/toplas/FosterGMPS07}
have proposed a domain specific language to define invertible transformations. Once a transformation is specified, the
inverse transformation is automatically derived. This enables them, for instance, to share data between several
applications that require different representations. Note that, a view can be partial and the original representation
can be required in order to transform a modified view back to its original form by injecting the updated data.
This work has been extended in order to deal with classes of
equivalent representations (\eg two lists of associations (key,value) can be equivalent even if the order of the pairs
are different) \cite{DBLP:conf/popl/BohannonFPPS08}.

%Partial views have been considered for specification matters
%(\emph{voir refs envoyees par Jean-Claude}). Here we handle the whole
%program in each view so that each view is ``independent'' of the
%others. 
%\fbox{REMI: je n'ai pas touche ici}

%More recently, work on lenses~\cite{Pierce} use the initial
%program to inject a view into an initial ``architecture''.
% 
%As the whole information is available in each view, it may be easier
%to implement evolutions in a view (otherwise, we could miss a
%propagation of the evolution in the missing part, in particular if a
%refactorer is used to implement the evolution (add a parameter to a
%function, add a constructor to the datatype...).
%
%\emph{(mais si le module est vraiment modulaire, la modification se limite au module, mais en pratique, le module n'est pas hermétique)}

Views have also been introduced at the specification level. For instance,
Jackson \cite{226249} shows how to compose Z specifications  on different
views of the same state. At the specification level, the composition is not
computational (it does not require transformation) but declarative: invariants
relate the different views of the state. 

Literate programming \cite{DBLP:journals/cj/Knuth84} proposes to invert the role of code and comments
(comments becomes the main view of a program and is commented by a few pieces of code). More
importantly, literate programming enables to decompose and reorder pieces of programs. This way,
the literate view for human has to be transformed into the code view for the compiler. However, the
transformation is single way. This does not help for the maintenance or the evolution problem 
(since the code is not transformed but only reordered) but this proves its is important to present 
alternative views for the programmer.

\bibliographystyle{abbrv}
%\bibliography{biblio}  %%%%%%%%%%%%%%%%%%%%%%%%%%%%%%%%%%%%%%%%%%%%%%%%%%

\begin{btSect}{websites}
\section*{Web Sites}
\btPrintCited
\end{btSect}

\begin{btSect}{biblio}
\section*{References}
\btPrintCited
\end{btSect}

\appendix

\section{Interfaces added to Refactoring Operations}
\label{app-added-interfaces}

\paragraph{Introduce new def.}%%%%%%%%%
\begin{itemize}

\item \texttt{haskell-refac-exhibitFunction f c n m} 

 In the equation
  concerning the constructor \texttt{c} in the definition of
  \texttt{f} in the module \texttt{m}, create a new local definition
  for the right hand side of the equation as \texttt{n} .

\item \texttt{haskell-refac-newDefFunApp f n f' m} 

 Find an
  application of the identifier \texttt{f} to \texttt{n} arguments in the module \texttt{m} and
  create a new local definition for that application as \texttt{f'}.

\end{itemize}

\paragraph{Generalise def.}%%%%%%%%%%%%%%%
\begin{itemize}

\item
\texttt{haskell-refac-generalise f c f' m n x curry "OtherType"}

In the module \texttt{m}, in the equation concerning the constructor \texttt{c} of the definition of \texttt{f}, let $v$ be the \texttt{n}$^{th}$ argument of the constructor \texttt{c}  in the pattern of the equation, generalise $v$ in the local definition of \texttt{f'} and name \texttt{x} the new argument.

The flag \texttt{curry} indicates whether the arguments of the constructor are curried or not to count the arguments.

\item
\texttt{haskell-refac-generalise f c f' m n x curry "RecType"}

In the module \texttt{m}, in the equation concerning the constructor \texttt{c} of the definition of {f}, let $v$ be the \texttt{n}$^{th}$ argument of the constructor \texttt{c}  in the pattern of the equation, generalise \emph{an application of \texttt{f} to $v$} in the local definition of \texttt{f'} and name \texttt{x} the new argument.

\item
\texttt{haskell-refac-generaliseIdent f m v x}

In the definition of \texttt{f} in the module \texttt{m}, generalise the variable \texttt{v} and name the new parameter \texttt{x}. 

\end{itemize}

\paragraph{Lift def to top level.}%%%%%%%%%%%%
\begin{itemize}

\item
\texttt{haskell-refac-makeGlobalOfLocalIn f d m}

Lift the definition of \texttt{d} at the top-level. \texttt{d} is declared inside the definition of \texttt{f} in the module \texttt{m},

\end{itemize}

\paragraph{Rename.}%%%%%%%%%%%
\begin{itemize}

\item
\texttt{haskell-refac-renameTopLevel f m f'}

Rename \texttt{t} declared at the top-level in the module \texttt{m} into \texttt{f'}.

\end{itemize}

\paragraph{Move def to another module.}%%%%%%%%%%%
\begin{itemize}

\item
\texttt{haskell-refac-moveDefBetweenModules f m m' }

Move the top-level definition of \texttt{f} from module \texttt{m} to module \texttt{m'}.

\end{itemize}

\paragraph{Unfold def./Fold def.}%%%%%%%%%%%%%%
\begin{itemize}

\item
\texttt{haskell-refac-unfoldInstanceIn d f m}

Replace an instance of the identifier \texttt{d} by the boy of its definition, in the body of \texttt{f} in module \texttt{m}. If possible, a beta-reduction is applied (see the corresponding HaRe operation).

\item
\texttt{haskell-refac-foldToplevelDefinition f m}

Fold the definition of function \texttt{f} of module \texttt{m} (see the corresponding HaRe operation).

\end{itemize}

\paragraph{Generative Fold}
\begin{itemize}

\item
\texttt{haskell-refac-generativeFold f "$i$" m}

Select an application of the identifier \texttt{f} to $i$ arguments in m and apply HaRe Generative Fold operation on it (a comment must be present before the affected declaration, see the HaRe operation).

\end{itemize}

\paragraph{Remove def.}%%%%%%%%%%%%
\begin{itemize}

\item
\texttt{haskell-refac-removeDefCmd f m}

Remove the definition of \texttt{f} in the module \texttt{m}. \texttt{f} must not be used elsewhere.

\item
\texttt{haskell-refac-removeLocalDef d f m}

Remove the definition of \texttt{d} which is local to \texttt{f} in the module \texttt{m}. \texttt{f} must not be used elsewhere.

\end{itemize}

\paragraph{Clean imports, Remove from export.}%%%%%%%%%%%%%%%
\begin{itemize}

\item
\texttt{haskell-refac-cleanImportsCmd m}

Call the Clean imports operation on the module \texttt{m} (remove the useless imports).

\item
\texttt{haskell-refac-rmFromExports f m}

Remove \texttt{f} from the explicit exports of the module \texttt{m}. \texttt{f} must not be used in an other module. 

\end{itemize}

\section{Added Refactoring Operations}%%%%%%%%%%%%%%%%%%%%%%%%%%%%%%%%%
\label{app-added-operations}

\begin{description}
%%%%%%%%%%%%%%%%%%%%%%%%%%%%%%%%%%%%%%%%%%%%%%%%%%%%%%%
\item[Simplify a \emph{case} pattern matching.] This operation transforms the first code below into the second code below.

\begin{lstlisting}
case (e1, e2, e3) of
         (y, p11, p12) -> b1
         (y, p21, p22) -> b2
\end{lstlisting}
\begin{lstlisting}
let y = e1
in case (e2, e3) of
            (p11, p12) -> b1
            (p21, p22) -> b2
\end{lstlisting}

   The patterns and the matched expression have to be tuples.

   The refactoring applies when there is a same identifier at the same position in all the pattern tuples.

   Input : select the whole case expression.

%   Comment 1 : this refactoring is purely structural/syntactical, there is no semantic analysis.

%   Comment 2 : This refactoring has not been tested nor proved correct, there is no guaranty it works.
%   Comment 3 : Unlike for Haskell Refactorer refactorings, this refactoring source code does not use StrategyLib. We have tried to avoid monads also (just because the author is not easy with it).

%Additional interface:
 \texttt{haskell-refac-simplifyCasePattern f m} 

Apply the above operation on a \emph{case} expression in the definition of \texttt{f} in the module \texttt{m}.

%%%%%%%%%%%%%%%%%%%%%%%%%%%%%%%%%%%%%%%%%%%%%%%%%%%%%%%%%%%%%%%%%%
\item[Transform a \emph{case} pattern matching into equations.] Transforms the first code below into the second code below.

\begin{lstlisting}
f x = case (x) of
            p1 -> e1
            p2 -> e2
\end{lstlisting}

\begin{lstlisting}
f p1 = e1
f p2 = e2
\end{lstlisting}

The pattern in the initial case has to be reduced to a variable which occurs as a parameter of the function.

A second version of this function is available for pattern matching on pairs:

\begin{lstlisting}
f x y = case (x,y) of
            (p11,p12) -> e1
            (p21,p22) -> e2
\end{lstlisting}

\begin{lstlisting}
f p11 p12 = e1
f p21 p22 = e2
\end{lstlisting}

\texttt{haskell-refac-caseToEq f m}\\
\texttt{haskell-refac-caseToEq2 f m}

Select the case expression which is at the top-level of the body of the declaration of \texttt{f} in the module \texttt{m} and transform it into a set of equations.

%%%%%%%%%%%%%%%%%%%%%%%%%%%%%%%%%%%%%%%%%%%%%%
\item[Copy a declaration into a comment.] Makes a copy of a declaration into a comment placed just above the declaration (to be used before applying Generative Fold).

\texttt{haskell-refac-duplicateIntoComment f m}
Copy of the declaration of \texttt{f} of the module \texttt{m}.

%%%%%%%%%%%%%%%%%%%%%%%%%%%%%%%%%%%%%%%%%%%%%%
\item[Remove a comment.] Deletes a comment.

\texttt{haskell-refac-rmCommentBefore f m}
Delete the comment occurring before the declaration of \texttt{f} at the top-level of the module \texttt{m}.

\end{description}

\end{document}